\documentclass[pre,twocolumn,amsmath,superscriptaddress,showpacs,amssymb,aps]{revtex4}

\usepackage{graphicx}% Include figure files

\begin{document}

\title{Episodic synchronization in dynamically driven neurons}

\author{Pablo Balenzuela}
\email{balen@df.uba.ar}
\affiliation{Departament de F\'isica i Enginyeria Nuclear, Universitat Polit\`ecnica de Catalunya, Colom 11,
E-08222 Terrassa, Spain}
\affiliation{Departamento de F\'isica, Facultad de Ciencias Exactas y Naturales,
Universidad de Buenos Aires, Pabell\'on 1, Ciudad Universitaria (1428), Buenos Aires,
Argentina}

\author{Javier M. Buld\'u}
%\email{jordi.g.ojalvo@upc.edu}
%\author{...}
\affiliation{Departament de F\'isica i Enginyeria Nuclear, Universitat Polit\`ecnica de Catalunya, Colom 11,
E-08222 Terrassa, Spain}
\affiliation{Nonlinear Dynamics and Chaos Group,
Departamento de  
Ciencias de la Naturaleza y F\'{\i}sica Aplicada, Universidad Rey Juan Carlos, Tulip\'an s/n,
28933 M\'ostoles, Madrid, Spain}

\author{Marcos Casanova}
%\email{jordi.g.ojalvo@upc.edu}
%\author{...}
\affiliation{Departament de F\'isica i Enginyeria Nuclear, Universitat Polit\`ecnica de Catalunya, Colom 11,
E-08222 Terrassa, Spain}

\author{Jordi Garc\'ia-Ojalvo}
\email{jordi.g.ojalvo@upc.edu}
%\author{...}
\affiliation{Departament de F\'isica i Enginyeria Nuclear, Universitat Polit\`ecnica de Catalunya, Colom 11,
E-08222 Terrassa, Spain}

\date{\today}

\begin{abstract}
We examine the response of type II excitable neurons to trains of synaptic pulses,
as a function of the pulse frequency and amplitude. We show that the resonant behavior characteristic of type II excitability, already described for harmonic inputs, is also present
for pulsed inputs. With this in mind, we study the response of
neurons to pulsed input trains whose frequency varies continuously in time, and
observe that the receiving neuron synchronizes episodically to the input pulses,
whenever the pulse frequency lies within the neuron's locking range.
We propose this behavior as a mechanism of rate-code detection in neuronal
populations. The results are obtained both in numerical simulations of the  
Morris-Lecar model and in an electronic implementation of the FitzHugh-Nagumo
system, evidencing the robustness of the phenomenon.  
\end{abstract}
\pacs{87.19.La, 05.45.Xt, 87.10.+e}

%\keywords{Suggested keywords}
%Use showkeys class option if keyword display desired

\maketitle

\section{\label{sec:intro}Introduction}

Neurons exhibit all-or-none responses to external input signals. The main
function of this thresholding behavior is to process information in a way that is efficient
and robust to noise \cite{lindner04}. Input signals received by most non-sensory
neurons take the form of pulse trains, coming from the spiking activity of
neighboring neurons. Therefore, in order to understand the mechanisms of
information processing in neural systems, it is very important to characterize in
detail the response of
neurons to pulse trains. Furthermore, realistic pulse trains are intrinsically
dynamical, with an instantaneous firing frequency that varies continuously in time.
It is therefore necessary to assess the influence of this non-stationarity in 
the neuronal response. This paper addresses these questions.

Most studies of driven neurons have been restricted so far to harmonic driving
signals \cite{lee99,yu01,parmananda02,laing03,sthilaire04}. Many of these
works have shown that for certain types of neurons, i.e. those exhibiting what is called
type II excitability, a resonant behavior arises with respect to the external driving
frequency. Excitability in those neurons is usually associated with an
Andronov-Hopf bifurcation, which leads to the existence of subthreshold oscillations
in the excitable regime. When the frequency of these oscillations equals that
of the harmonic driving, a resonance arises.

It is to be expected that a similar resonant behavior exists for pulsed inputs.
In that case, the same pulse train impinging on two different neurons could
elicit a response on only one of them, i.e. on the one that is tuned to resonate
with the incoming pulse frequency. This behavior has indeed been observed
experimentally in a rat's neocortical pyramidal neuron that innervates another
pyramidal neuron and an interneuron; a bursting input from the innervating
neuron produced an action potential in the interneuron but not in the
second pyramidal neuron \cite{markram98}. This behavior was interpreted
in terms of a differential frequency-dependent facilitation and depression,
respectively, and as such was studied by Izhikevich and co-workers
\cite{izhikevich02,izhikevich03}.
Here we propose a simpler mechanism for this phenomenon,
relying only on the resonant behavior of the processing neuron.
This mechanism could provide a means for
distinguishing between firing rates, whose controlled variation lies at the heart of
the rate coding approach to information processing by neurons.

Our results show that type II excitable neurons exhibit a resonant response
with respect to the frequency of input pulse trains. This behavior leads to
episodic synchronization between the neuron's output and an input with
dynamically varying firing rate. Episodic synchronization has previously been
reported in coupled lasers with intrinsic dynamics \cite{buldu06}. Here
we extend that property to externally driven excitable systems. Two types
of systems have been investigated: a Morris-Lecar model (Sec.~\ref{sec:ml})
and an electronic implementation of the FitzHugh-Nagumo model
(Sec.~\ref{sec:elect}).

\section{Morris-Lecar model}
\label{sec:ml}

\subsection{Model description}

We consider neurons whose dynamical behavior is described by the Morris-Lecar model \cite{morris81},
\begin{eqnarray}
\frac{dV}{dt} & = & \frac{1}{C_m}(I_{\rm app} - I_{\rm ion}-I_{\rm syn})+D\xi(t) \label{eqV} \\
\frac{dW}{dt} & = & \phi \Lambda(V)[W_{\infty}(V) - W]   \label{eqW}
\end{eqnarray}
where $V$ and $W$ represent the membrane potential and the fraction of 
open potasium channels, respectively. $C_m$ is the membrane capacitance per unit area and $\phi$ is the decay rate of $W$. The neuron is affected by several
currents, including an external current $I_{\rm app}$, a synaptic current
$I_{\rm syn}$, and an ionic current given by
\begin{eqnarray}
&&I_{\rm ion}  =  g_{Ca}M_{\infty}(V)(V-V^0_{Ca})+ \nonumber \\
 & & \qquad\qquad\qquad g_KW(V-V^0_K)+g_L(V-V^0_L)\,.
\label{eqIion}
\end{eqnarray}
In this expression, $g_a$ ($a=Ca,K,L$) are the conductances and $V^0_a$ the
resting potentials of the calcium, potassium and leaking channels, respectively. We
define the following functions of the membrane potential:
\begin{eqnarray}
&& M_{\infty}(V) =  \frac{1}{2}\left[1+\tanh\left(\frac{V-V_{M1}}{V_{M2}}\right)\right]      \\
&&W_{\infty}(V) = \frac{1}{2}\left[1+\tanh\left(\frac{V-V_{W1}}{V_{W2}}\right)\right]      \\
&&\Lambda(V) = \cosh\left(\frac{V-V_{W1}}{2V_{W2}}\right)\,,
\end{eqnarray}
where $V_{M1}$, $V_{M2}$, $V_{W1}$ and $V_{W2}$ are constants to be specified later. The last term in Eq.
(\ref{eqV}) is a white Gaussian noise term of zero mean and amplitude $D$.

In the absence of noise, an isolated Morris-Lecar neuron shows a bifurcation to a limit cycle for increasing applied current $I_{\rm app}$ \cite{sthilaire04}. Depending on the parameters, this bifurcation can be of the saddle-node or the 
subcritical Hopf types, corresponding to either type I or type II excitability,
respectively. 
%We chose this last option for the numerical calculations presented in this paper. 
The specific values of the parameters used are shown in table \ref{tab:ML} \cite{tsumoto06}. For these parameters, the threshold values of the applied current
under constant stimulation are $39.7$~mA for type I and $46.8$~mA for type II.

In this paper we analyze the behavior of a neuron driven by a synaptic current. To
that end, we use the simplified model of chemical synapse proposed in Ref.
\cite{destexhe}, according to which the synaptic current is given by
\begin{equation}
I_{\rm syn} = g_{\rm syn}r(t)(V-E_s), \label{syn}
\end{equation}
where $g_{\rm syn}$ is the conductance of the 
synaptic channel, $r(t)$ represents the fraction of bound receptors,
and $E_s$ is a parameter whose value determines the type of synapse: if $E_s$
is larger than the rest potential the synapse is excitatory, if smaller it is inhibitory;
here we consider an excitatory synapse with $E_s=0$~mV.
The fraction of bound receptors, $r(t)$, follows the equation
\begin{equation}
\label{receptor}
\frac{dr}{dt}= \alpha [T](1-r) - \beta r\,,
\end{equation}
where $[T]=T_{max} \theta(T_0 +\tau_{\rm syn}-t)\theta(t-T_0)$ is the concentration
of neurotransmitter released into the synaptic cleft by the presynaptic neuron, whose
dynamics is also given by Eqs.~(\ref{eqV})-(\ref{eqW}) with no synaptic input.
$\alpha$ and $\beta$ are
rise and decay time constants, respectively, and $T_0$ is the
time at which the presynaptic neuron fires, which happens whenever the presynaptic membrane potential exceeds 
a predetermined threshold value, in our case chosen to be $10$~mV. This thresholding mechanism lies at the origin 
of the nonlinear character of the synaptic coupling. 
The time during which the synaptic connection is active is roughly given by $\tau_{\rm
syn}$. The values of the coupling parameters that we use \cite{destexhe} are specified in Table \ref{tab:ML}. The equations were integrated using the Heun method \cite{nises}, which is a second order Runge-Kutta algorithm for stochastic equations.

\begin{table}[htbp]
\begin{center}
\begin{minipage}[t]{2.5in}
\centering
\begin{tabular}{|c|c|}
 \hline
  \textbf{Parameter}  & \textbf{Morris-Lecar TII (TI)} \\ \hline\hline
 $C_m$  &  $5\,\mu \mathrm{F/cm}^2$    \\ \hline
 $g_K$  &  $8\,\mu \mathrm{S/cm}^2$   \\  \hline
 $g_L$  &  $2\,\mu \mathrm{S/cm}^2$     \\ \hline
 $g_{Ca}$  & $4.0\,\mu \mathrm{S/cm}^2$    \\   \hline
 $V_K$  &  $-80\,\mathrm{mV}$     \\ \hline
 $V_L$  &  $-60\,\mathrm{mV}$     \\ \hline
 $V_{Ca}$ & $120\,\mathrm{mV}$     \\ \hline
 $V_{M1}$  &  $-1.2\,\mathrm{mV}$     \\ \hline
 $V_{M2}$  &  $18\,\mathrm{mV}$     \\ \hline
 $V_{W1}$  &  $2\,\mathrm{mV}$ ($12\,\mathrm{mV}$)    \\ \hline
 $V_{W2}$  &  $17.4\,\mathrm{mV}$     \\ \hline
 $\phi$  & $1/15\, \mathrm{s}^{-1}$     \\ \hline
 \end{tabular}
 \begin{tabular}{|c|c|}
 \hline
 \textbf{Parameter}  & \textbf{Synapse} \\ \hline\hline
  $\alpha$  & $2.0\, \mathrm{ms}^{-1} \mathrm{mM}^{-1}$ 	\\ \hline
 $\beta$   & $1.0\, \mathrm{ms}^{-1}$ 	\\ \hline
 $T_{max}$   & $1.0\, \mathrm{mM}$ 	\\ \hline
 $g_{\rm syn}$ & (specified in each case)	\\ \hline
 $\tau_{\rm syn}$ & $1.5\, \mathrm{ms}$  \\ \hline
 $E_s$ & $0\,\mathrm{mV}$	\\ \hline
\end{tabular}
\caption{Parameters values of the Morris-Lecar and synapse models used in this work.
\label{tab:ML}}
\end{minipage}
\end{center}
\end{table}
		
\subsection{Response diagram of a periodically driven driven neuron}

First we analyze how a Morris-Lecar neuron responds to periodic inputs of varying
frequencies. Specifically, we ask how large the signal needs to be in order to
elicit spikes in the receiving neuron. It is also important to characterize the frequency
of  spiking in terms of frequency of the input.
As mentioned in the introduction, this question has already been addressed, in the
case of harmonic inputs, for different neuronal models, including the Morris-Lecar
model \cite{chinosML}. We will now compare these results with those obtained for
a pulsed input.
For the Morris-Lecar model, one can expect a completely different behavior between
the type I and the type II cases, given that the bifurcation to a limit cycle is a
saddle-node bifurcation in the former case and a Hopf bifurcation (with the well
known eigenfrequency associated to the spiral fixed point) in the latter.

\begin{figure}[htb]
\includegraphics[height=9cm,angle=270,keepaspectratio,clip]{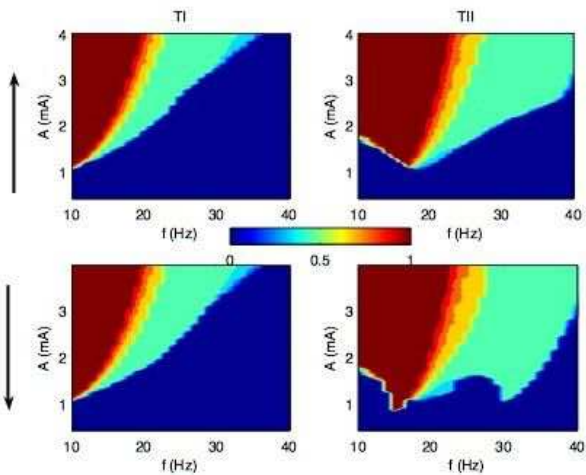}
\caption{\label{fig1} Response diagram of Morris Lecar neurons for
a harmonic input: $f_{\rm out}/f_{\rm in}$ is plotted (color-coded) as a function of
the amplitude and frequency of the applied current. 
Left plots: type I neuron (with $I_0=39$~mA), right plots: type II neuron
(with $I_0=46$~mA). The response is measured for increasing $A$ in the upper plots,
and for decreasing $A$ in the lower plots.
}
\end{figure}

We first consider an isolated neuron without synaptic inputs ($I_{\rm syn}=0$),
but subject to a harmonic modulation of the applied current $I_{\rm app}$ with the
form,
\begin{equation}
I_{\rm app} =  I_0+A \cos(2\pi f_{\rm in}t)\,.
\label{eq:arm}
\end{equation}
In order to quantify the response of the neuron to this harmonic input,
we plot in Fig.~\ref{fig1} 
(in color scale) the ratio between the output and the input frequencies,
$f_{\rm out}/f_{\rm in}$, as a function of the amplitude $A$ and frequency $f_{\rm in}$
of the applied current (\ref{eq:arm}). The figure compares the
resulting  {\it response diagrams}
of type I and type II neurons. 
To obtain these plots, $A$ is varied for fixed $f$, while
using as initial condition for a given $A$ the final state of the previous $A$ value.
In the upper panels $A$ increases, thus showing the stability of the rest state, while
in the lower panels $A$ decreases, this indicating the stability of the limit cycle.
The figure
shows that type II neurons have a region of bistability, where the fixed point and
the limit cycle coexist. In contrast, for type I neurons the two plots are basically
the same, indicating and absence of bistability.
 
There is another qualitative difference between types I and II that can be observed
in Fig.~\ref{fig1}. In the type I neuron, the critical modulation
amplitude for spiking increases monotonically with the frequency of the stimulus.
On the other hand, in the type II neuron the critical amplitude exhibits
a minimum for a given nonzero frequency, in our case around $20$~Hz.
This behavior can be understood as resulting from the 
subthreshold damped oscillations characteristic of type II excitability \cite{sthilaire04}.

We now characterize the response of a neuron to an input train of periodic synaptic
pulses of varying frequencies and amplitudes. To that end, we drive the neuron with a
synaptic current with the form
\begin{equation}
I_{\rm syn} =  A \,r(t,f_{\rm in})(V(t)-E_s)\,.
\label{syn1}
\end{equation}
In this model, the frequency is given by the dynamics of $r(t,f_{\rm in})$, described in Eq.~(\ref{receptor}), assuming that the presynaptic firings
occur periodically with frequency $f_{\rm in}$. 
In this way, we can quantify the response of the neuron in terms of the efficiency in
responding to a periodic synaptic input with a given frequency and amplitude, as shown in the harmonic case.
\begin{figure}[htb]
\includegraphics[height=9cm,angle=270,keepaspectratio,clip]{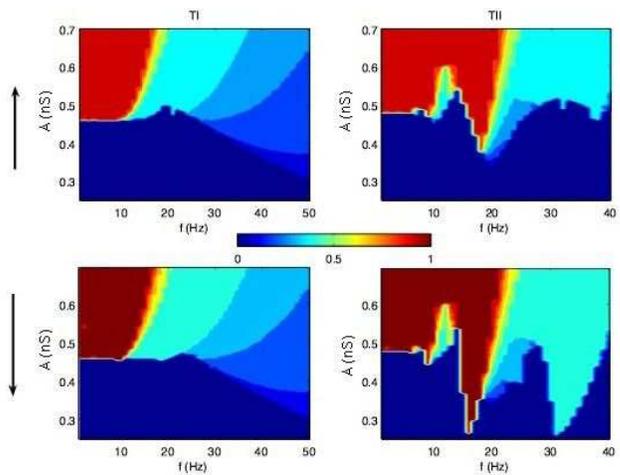}
\caption{\label{fig2} Response diagrams of Morris Lecar neurons to a periodic synaptic
(pulsed) input: $f_{\rm out}/f_{\rm in}$ is plotted (in color scale) as a function of the
amplitude and the frequency of the  synaptic current.
Left plots: type I neuron (with $I_{\rm app}=39$~mA), right plots: type II neuron (with
$I_{\rm app}=46$~mA). The response is measured for increasing $A$ in the upper plots,
and for decreasing $A$ in the lower plots. 
}
\end{figure}
Figure \ref{fig2} shows the corresponding response diagrams,
i.e. $f_{\rm out}/f_{\rm in}$ as a function of $A$ and $f_{\rm in}$,
for both excitability types and for increasing (top) and decreasing (bottom) $A$. 
The behavior shows 
features common to the harmonic case, such as the existence of the same resonant
frequency in type II for both kinds of inputs. But there are also very interesting 
differences between them, specially in the high and zero frequency limits.

The main difference between the harmonic and pulsed input cases is the approach
to the DC threshold current ($39.7$~mA for type I and $46.8$~mA for type II).
While in the former case this happens for frequencies approaching zero (type I)
or resonance (type II), for pulsed inputs it happens for high frequencies, i.e. when
the signal period is of the order of the pulse width. This is the reason for the
appearance of a spiking region 
at high frequencies for pulsed inputs, which is absent in the harmonic case
\cite{DC}.  
Also, in the low-frequency limit one can observe, for pulsed inputs, a constant value
of the critical amplitude. This is related with the fact that, when the input period is
high enough
with respect to the pulse width, the system response is essentially independent of
the period.

%A comparison of Figs.~\ref{fig1} and \ref{fig2} shows that
%while in the harmonic case the approach to the DC threshold current
%($39.7$~mA for type I and $46.8$~mA for type II)
%occurs for low frequencies,
%for pulsed inputs that limit is approached for high frequencies, and takes place
%when the signal period is of the order of the pulse width (which is given by
%$\tau_{\rm syn}$). This is the reason for the appearance of a spiking region 
%for high frequencies when the input is pulsed, which is absent in the harmonic case.  

\subsection{The dynamical case: Variation of the input frequency}

In the previous section, we have characterized the behavior of a neuron subject to a
pulsed synaptic current of fixed frequency. Our results show that type II neurons exhibit a resonant behavior, defined by the existence of an optimal frequency for which the critical amplitude for spiking is minimal. The question now is, what happens if the frequency of the input train varies dynamically, which is a more realistic situation for a non-sensory neuron. 

To answer this question, we made simulations with two Morris-Lecar neurons
coupled unidirectionally through a chemical synapse. The input neuron operates
in the limit cycle regime and is considered to be type I, so that 
we can control its spiking frequency by varying its applied
current $I_{\rm app}$ \cite{sthilaire04}. This neuron is 
synaptically coupled to a type II neuron operating in an excitable regime,
with a coupling strength $g_{\rm syn}$ such that the receiving neuron only fires in a
given range of frequencies (i.e. the coupling is such that the amplitude of the
input pulses lies below the critical amplitude at zero frequency but above its
minimum at resonance; this
corresponds e.g. to a horizontal line at around 0.4 mV in the right plots of
Fig.~\ref{fig2}).

\begin{figure}[htb]
\includegraphics[height=6cm,keepaspectratio,clip]{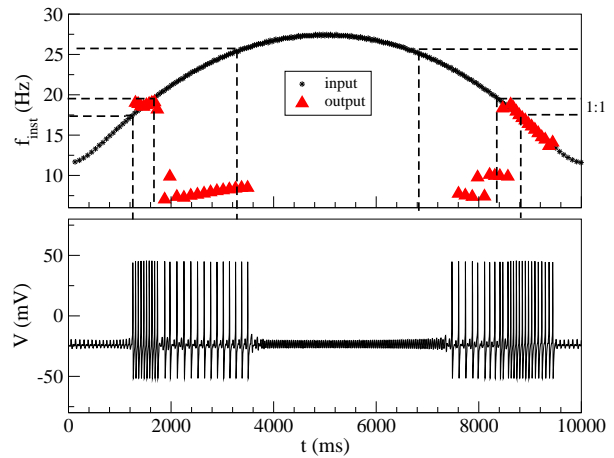}
\caption{\label{fig4}Upper plot: instantaneous frequency versus time of the input
neuron (black stars) and of the receiving neuron (red triangles). The three
horizontal dashed lines indicate the boundaries of the locking region for
the selected synaptic strength ($g_{\rm syn}=0.43$~nS); the 1:1 region
is specifically shown, and labeled at the right. Lower plot: time series of the
receiving neuron.
}
\end{figure}

Figure \ref{fig4} shows what happens when the firing frequency of the input neuron
first increases and then decreases in the range 13-28~Hz. The plot compares
the instantaneous firing frequency of both neurons, relating them with the boundaries
of the locking range of the second neuron, indicated by horizontal dashed lines;
the 1:1 locking region is specifically shown.
It can be seen that as the input
frequency increases (first half of the plot), the receiving neuron
starts spiking with approximately 1:1 frequency ratio when the input frequency falls
within the corresponding locking range, the ratio decreasing when the input frequency
exceeds $\sim$20~Hz. Spiking persists while the input frequency remains in
the wider (not 1:1) locking range, and is maintained even for a while after the input
finally exits the locking region. A similar behavior is observed for decreasing
frequencies, but the ``inertia'' observed at the exit of the locking region is larger
than for increasing frequencies. The time series of the receiving neuron is shown
in the lower plot; the episodes of synchronization with the input signal are clearly
observed.

\begin{figure}[htb]
\includegraphics[height=7cm,keepaspectratio,clip]{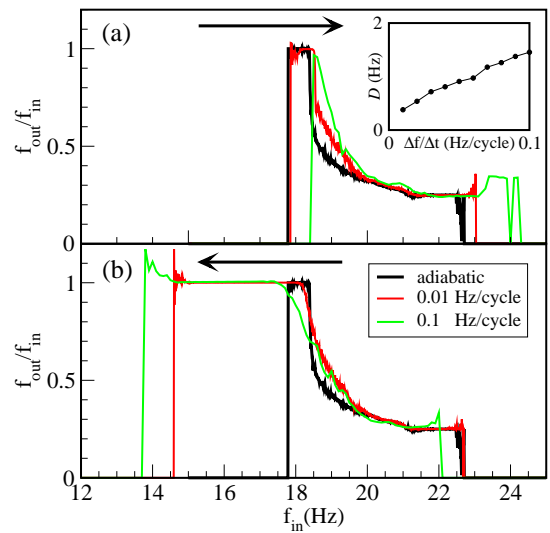}
\caption{\label{fig6}Response of a type II neuron to a controlled variation
of the input frequency, for two different rates of this variation,
$f'=\Delta f/\Delta t$,  for $g_{\rm syn}=0.38$~nS,  compared with the adiabatic response (which is a horizontal cut of the {\rm response diagram} at $A=0.38$~nS).
}
\end{figure}

In order to understand the dynamic driving effects reported above, and particularly
the locking asymmetry observed between an increase and a decrease in the
input firing frequency, we now study the neuron response to a controlled
variation of the frequency for different variation rates. To that end, we consider a
synaptic input whose frequency is 
uniformly changing from one cycle to the next at a rate $f'=\Delta f/\Delta t$, and measure the response of the neuron in terms of its instantaneous output frequency
$f_{\rm out}$. Figure~\ref{fig6} shows the response diagram for a fixed synaptic
strength and different frequency variation rates,
for both increasing and decreasing input frequencies $f_{\rm in}$.
The figure shows that the slower the variation rate, the closer the response is to an adiabatic passage, as expected. Additionally, the results indicate that the
persistence of the output neuron in the firing state (even when the input signal
has left the locking region) is much larger when the frequency
decreases than when it increases. This is consistent with the asymmetric
response exhibited in Fig.~\ref{fig4}, and can be expected to arise from the
asymmetric shape of the response function $f_{\rm out}/f_{\rm in}$, which
is equal to 1 for small frequencies and moderately smaller than 1 for large
frequencies. Evidently the neuron prefers to respond in a 1:1 regime, which
produces a larger persistence for decreasing frequencies.

To further quantify the approach to the adiabatic response in terms of the rate of
change in the input frequency, we can define a distance ${\cal D}$ to this adiabatic
response as the absolute value of the difference between the area of the neuron's
response diagram at a given rate $f'$, as plotted in Fig.~\ref{fig6}, 
and the area of the adiabatic response:
\begin{equation}
{\cal D}=\int_{f_{\rm min}}^{f_{\rm max}} \left[\left(\frac{f_{\rm in}}{f_{\rm out}}
\right)_{f'} - \left(\frac{f_{\rm in}}{f_{\rm out}}\right)_{\rm adiab}\right] df_{\rm in}\,.
\end{equation} 
This measure is plotted in the inset of Fig.~\ref{fig6} as a function of the rate of
change in the input frequency. The plot shows that the distance increases as the
frequency changes more rapidly, as expected.

\begin{figure}[htb]
\includegraphics[height=6cm,keepaspectratio,clip]{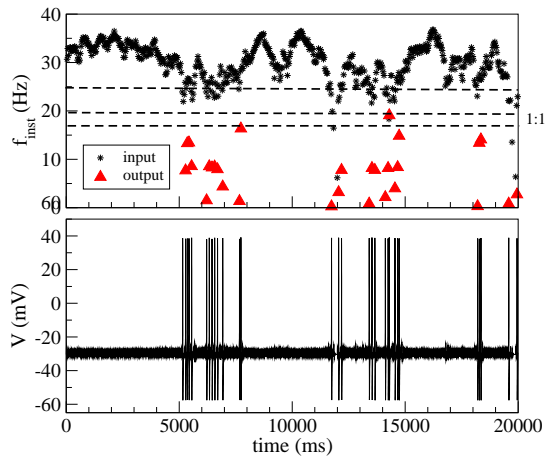}
\caption{\label{fig5}Upper panel: Instantaneous frequency of the synaptic input (black stars) and of the
output neuron (red triangles) as a function of time. The horizontal dashed lines
delimit the region of locking
according to the adiabatic calculations 
of Fig.~\protect\ref{fig2} for the synaptic strength used
($g_{\rm syn}=0.38$~nS). Lower panel: time series of the input signal.
}
\end{figure}

The dynamical response described in the previous paragraphs leads to 
episodic synchronization when the input pulse train exhibits a varying
firing rate. This situation is shown in Fig.~\ref{fig5}, in which an input
pulse train whose firing rate takes the form, by way of example,
of an Ornstein-Uhlenbeck noise with amplitude $A_{\rm ou}=0.1$~mA and
correlation time 
$\tau_{\rm ou}=1$~s in the $\sim 20-36$~Hz range. The response of the 
second neuron for $g_{\rm syn}=0.38$~nS is displayed in the bottom plot, and
exhibits clear episodes of synchronization with the input signal, whenever the
firing rate of the later falls (approximately) within the locking range of the
neuron for the coupling strength chosen (represented by horizontal dashed
lines in the figure). In that way the receiving neuron acts as a bandpass filter
for input pulse trains.

\section{FitzHugh-Nagumo circuit}
\label{sec:elect}

In order to show that the behavior reported in the previous Section is generic and
robust, we have reproduced the results with an electronic neuron, specifically with
an electronic implementation of the FitzHugh-Nagumo (FHN) model \cite{nagumo64}.
The circuit has been previously described in \cite{tor03}, where synchronization
between two FHN neurons was studied. A detailed description of the circuit can be
found at \cite{circuit}.  In our particular setup, a FHN neuron is excited
by a pulsed input of variable frequency.

\begin{figure}[htb]
\includegraphics[width=8cm,keepaspectratio,clip]{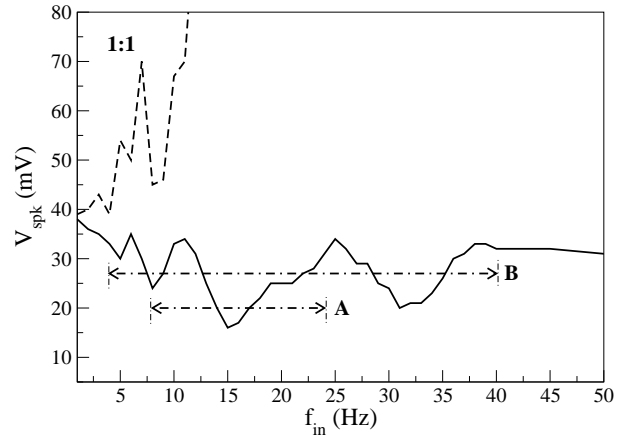}
\caption{\label{fig:exp_res_diag} Response diagram of an electronic FHN neuron
for a periodic input of pulses with $10$~ms width. The limit of the
region of 1:1 resonance is marked with a dashed line. 
}
\end{figure}

Following the procedure of Sec.~\ref{sec:ml}, we first 
determine the response of the electronic neuron to a train of periodic pulsed
inputs of fixed frequency. The pulses have the form of square pulses of $10$~ms
width. Figure~\ref{fig:exp_res_diag} shows the corresponding response diagram,
obtained by increasing the amplitude of the input pulses until the
neuron starts firing. Similar results (not shown here)
are obtained with pulses of different width.
At first glance, we can observe a resonance minimum around $f_{\rm in}=15$~Hz,
which confirms
that the FHN neuron is of type II. Two local minima are also observed around $f_{\rm in}=7.5$~Hz
and $f_{\rm in}=31$~Hz. It is worth noting that despite the spike threshold is low,
moderately large values of the input voltage are required to induce spiking at the
input frequency (see region 1:1 in Fig.~\ref{fig:exp_res_diag}).
 
\begin{figure}[htb]
\includegraphics[width=8cm,keepaspectratio,clip]{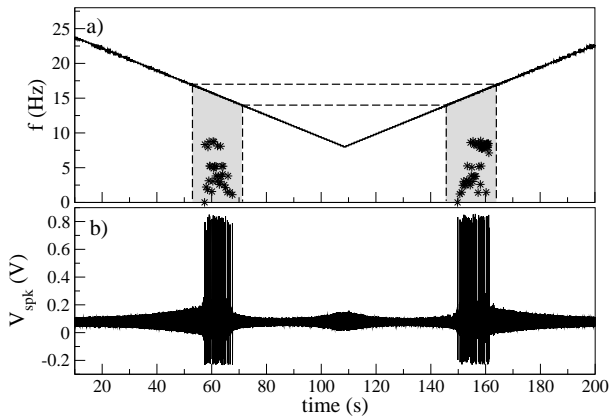}
\caption{\label{fig:exp_a} Response of a FHN electronic neuron to a pulsed
input of variable frequency. The input voltage corresponds to the value
marked by A in  Fig.~\ref{fig:exp_res_diag}.
Plot (a) shows the instantaneous frequency of the input pulses (solid line)
and of the neuron's output (stars). The shaded region corresponds to the frequency
ranges for which locking should occur. Plot (b) displays the time evolution of the
membrane voltage of the electronic neuron, which corresponds with the voltage U2 at condenser C1, in the circuit given in  \cite{circuit}. 
}
\end{figure}

\begin{figure}[htb]
\includegraphics[width=8cm,keepaspectratio,clip]{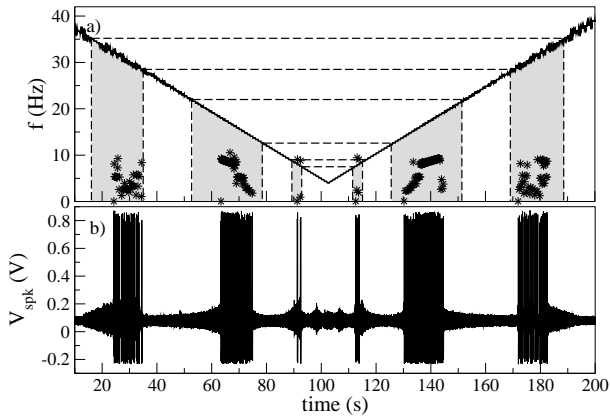}
\caption{\label{fig:exp_b}
Response of a FHN electronic neuron to a pulsed
input of variable frequency. The input voltage corresponds to the value
marked by B in  Fig.~\protect\ref{fig:exp_res_diag}. Figure layout is as described
in the caption of Fig.~\protect\ref{fig:exp_a}.
}
\end{figure}

We have thus a type II electronic neuron that exhibits a resonance at a frequency
close to $15$~Hz. Following again the approach of Sec.~\ref{sec:ml}, we now
subject the circuit to pulse trains with time-varying frequency. 
Specifically, the pulse frequency is made to depend linearly with time
(with both positive and negative slope). Similar results (not shown here)
are obtained with sinusoidal variations. The neuron response to this dynamical
input is shown in Figs.~\ref{fig:exp_a} and \ref{fig:exp_b},
where two different input voltages, corresponding to the values
denoted as A and B in Fig.~\ref{fig:exp_res_diag}, have been applied. 

In the case
of Fig. \ref{fig:exp_a}, the input signal scans the region marked with $A$ in 
Fig.~\ref{fig:exp_res_diag}. The upper plot shows
the instantaneous frequency of the input train (solid line) and of the FHN neuron
(stars). Figure~\ref{fig:exp_a}(b) plots the neuron's output. The results
show that the neuron pulses when the input frequency lies within the resonance
regions given in Fig.~\ref{fig:exp_res_diag}, and highlighted in gray in
Fig.~\ref{fig:exp_a}(a). This behavior is an agreement with the observations
made in the Morris-Lecar model. The fact that no inertia effects are seen when
the input frequency sweeps past the resonance region is due to the frequency
variation rate being very slow with respect to the characteristic time scales of
the system (adiabatic limit).

Figure~\ref{fig:exp_b} shows the system's behavior for a different value of the
input voltage (marked as B in Fig.~\ref{fig:exp_res_diag}), for which the input
frequency encounters three resonance
regions as it varies. Accordingly, the electronic neuron fires whenever
the input frequency lies inside any these regions, exhibiting clear episodes
of synchronization. In other words, the neuron
acts as a band-pass filter, with a frequency range that depends on the input
voltage level according to its response diagram.  

\section{Discussion}

Neurons are information-processing devices. The nature of coding in neuronal systems
is still an open question. One of the most favored views in the field is that
of rate coding, where the intensity of a signal is encoded in the firing rate. Neuronal
systems must therefore be able to distinguish between firing rates. We have proposed
a way to accomplish that, through the resonant behavior exhibited by type II neurons.
A population of neurons with different tuning characteristics, and therefore
distinct locking ranges, should be able to distinguish between different incoming
pulse frequencies by activating selectively different subpopulations 
that respond selectively to different frequencies.

We have systematically
analyzed the response of type I and II neurons to pulsed driving, compared it
with the standard case of sinusoidal driving, and observed the resonant behavior
of type II neurons. This phenomenology leads to episodic synchronization
between the input and the output of the neuron, when the input consists of
a train of pulses with dynamically varying frequency. The phenomenon has
been reported both in numerical simulations of the Morris-Lecar model, and in
an experimental implementation of the FitzHugh-Nagumo circuit. We expect this type
of behavior to underlie information processing in rate coding and decoding
neuronal populations in more complex brain networks.
 
\begin{acknowledgments}
We acknowledge financial support from MCyT-FEDER (Spain, project BFM2003-07850), and from the Generalitat de Catalunya. P.B. acknowledges financial support from the Fundaci\'on Antorchas (Argentina), and from a C-RED grant of the Generalitat de Catalunya.
\end{acknowledgments}


\begin{thebibliography}{99}
\bibitem{DC} In top left panel of  Fig.~\ref{fig2}, the spiking region stretches down below
$A=0.3$~nS, a lower value than expected. This happens because the amplitude
of the synaptic pulses is larger than $A$, due to modulation by the term $(V-E_s)$
shown in Eq.~(\ref{syn1}).
\end{thebibliography}
\end{document}